\documentclass[conference]{IEEEtran}

\usepackage[pdftex]{graphicx}
\usepackage{color}
\usepackage{comment}
\usepackage{numprint}
\usepackage{url}
\usepackage{tabularx}
\usepackage{tikz}

\newif\ifdraft

\ifdraft
  \definecolor{darkgreen}{rgb}{0,0.5,0}
  \newcommand{\woz}[1]{ {\noindent \textcolor{darkgreen} { Wozniak: #1 }}}
  \definecolor{darkblue}{rgb}{0,0,0.7}
  \newcommand{\katznote}[1]{ {\textcolor{darkblue}   { ***Dan:   #1 }}}
  \newcommand{\mike}[1]{ {\textcolor{red} { Mike: #1 }}}
  \definecolor{orange}{rgb}{0.7,0.5,0.0}
  \newcommand{\ian}[1]{{\textcolor{orange}{ Ian: #1 }}}
  \newcommand{\rusty}[1]{ {\textcolor{blue} { Rusty: #1 }}}
  \newcommand{\zhaonote}[1]{ {\textcolor{cyan} { Zhao: #1 }}}
  \newcommand{\skriedernote}[1]{ {\textcolor{cyan} { Scott: #1 }}}
  \newcommand{\ketanote}[1]{ {\textcolor{magenta} { Ketan: #1 }}}
  \definecolor{brown}{rgb}{0.4,0.2,0.0}
  \newcommand{\tim}[1]  { {\textcolor{brown}    { Tim: #1   }}}
  \newcommand{\TODO}{$\textcolor{red}{\pmb{\star}}$\xspace}
\else
  \newcommand{\katznote}[1]{}
  \newcommand{\woz}[1]{}
  \newcommand{\mike}[1]{}
  \newcommand{\ian}[1]{}
  \newcommand{\rusty}[1]{}
  \newcommand{\zhaonote}[1]{}
  \newcommand{\skriedernote}[1]{}
  \newcommand{\ketanote}[1]{}
  \newcommand{\tim}[1]{}
  \newcommand{\TODO}[1]{}
\fi

\definecolor{teal}{rgb}{0.06,0.3,0.3}
\definecolor{maroon}{rgb}{0.5,0.0,0.25}
\definecolor{darkblue}{rgb}{0.0,0.2,0.75}
\definecolor{darkred}{rgb}{0.7,0.0,0.0}
\definecolor{darkgreen2}{rgb}{0,0.35,0}

\definecolor{swiftbuiltincolor}{rgb}{0,0,0}
\definecolor{swiftstringcolor}{rgb}{0,0,0}
\definecolor{swiftcommentcolor}{rgb}{0,0,0}

\usepackage{eso-pic}
\newcommand\AtPageUpperMyright[1]{\AtPageUpperLeft{
 \put(\LenToUnit{0.5\paperwidth},\LenToUnit{-1cm}){
     \parbox{0.5\textwidth}{\raggedleft\fontsize{9}{11}\selectfont #1}}
 }}
\newcommand{\conf}[1]{
\AddToShipoutPictureBG*{
\AtPageUpperMyright{#1}
}
}

\conf{2015 IEEE International Conference on Cluster Computing \url{https://doi.org/10.1109/CLUSTER.2015.74}} 

\begin{document}

\title{Toward Interlanguage Parallel Scripting for \\
        Distributed-Memory Scientific Computing}

\author{\IEEEauthorblockN{
Justin M. Wozniak,\IEEEauthorrefmark{1}\IEEEauthorrefmark{3}
Timothy G. Armstrong,\IEEEauthorrefmark{2}
Ketan C. Maheshwari,\IEEEauthorrefmark{1} \\
Daniel S. Katz,\IEEEauthorrefmark{3}
Michael Wilde,\IEEEauthorrefmark{1}\IEEEauthorrefmark{3}
Ian T. Foster\IEEEauthorrefmark{1}\IEEEauthorrefmark{2}\IEEEauthorrefmark{3}}

  \IEEEauthorblockA{\IEEEauthorrefmark{1}
    Mathematics and Computer Science Division,
    Argonne National Laboratory,
    Argonne, IL, USA}
  \IEEEauthorblockA{\IEEEauthorrefmark{2}
    Dept. of Computer Science,
    University of Chicago,
    Chicago, IL, USA}
  \IEEEauthorblockA{\IEEEauthorrefmark{3}
    Computation Institute,
    University of Chicago and Argonne National Laboratory,
    Chicago, IL, USA}
}

\thispagestyle{empty}
\pagestyle{plain}

\maketitle

\begin{abstract}

Scripting languages such as Python and R have been widely adopted as
tools for the productive development of scientific software because of
the power and expressiveness of the languages and available
libraries. However, deploying scripted applications on large-scale
parallel computer systems such as the IBM Blue Gene/Q or Cray XE6 is a
challenge because of issues including operating system limitations,
interoperability challenges, parallel filesystem overheads due to the
small file system accesses common in scripted approaches, and other
issues. We present here a new approach to these problems in which the
Swift scripting system is used to integrate high-level scripts written
in Python, R, and Tcl, with native code developed in C, C++, and
Fortran, by linking Swift to the library interfaces to the script
interpreters.  In this approach, Swift handles data management,
movement, and marshaling among distributed-memory processes without
direct user manipulation of low-level communication libraries such as
MPI.  We present a technique to efficiently launch scripted
applications on large-scale supercomputers using a hierarchical
programming model.

\end{abstract}

\section{Introduction}

An increasing number of modern scientific applications and tools are
built by using a variety of languages and libraries. These complex
software products combine performance-critical libraries implemented
in native code (C, C++, Fortran) with high-level functionality
expressed in rapidly developed and modified scripts. Additional
specialized language-specific features may be used for concurrency,
I/O, the use of accelerators, and so on.  These development techniques
have been used in a wide range of application domains, from materials
science and protein analysis to power grid simulation.

Such applications and tools are commonly developed with the following
software development pattern. First, a native code library is built or
repurposed for the core processing.  Second, a collection of scripts
is built up around the core library or program to express the complex,
often dynamic, but less performance-critical coordination logic.  Such
``wrapper scripts'' may be developed with shell, Python, Tcl, or other
tools.  Third, when additional scalability is required, native code or
additional scripts are developed to deploy the application in some
distributed computing model such as MPI, Swift, some other grid
workflow system, or with custom wrapper scripts that submit jobs to a
scheduler such as PBS.

Swift~\cite{Swift_2011} is a programming language and runtime designed
to ease the software development methodology described above. Swift
has a well-defined concept of wrapper scripts, the ability to
coordinate calls to tools through its programming model, and built-in
support for many schedulers and data movement protocols.  The latest
implementation, Swift/T~\cite{Swift_T_2013}, generates an MPI program
from the Swift script and provides tools to run that program on
various scheduled resources. This approach has allowed Swift/T to
scale the execution of scripted applications to hundreds of thousands
of cores~\cite{STC_2014}.

The Swift/T framework supports direct calls to native code through
library loading and access. As described above, however, modern
scientific applications are built not only with native code, but also
with scripts and scripting interfaces to core libraries. Thus, to ease
the coordination of calls to tools in the Swift programming model, we
wish to support direct calls to script code without calling external
programs or forcing the user to master complex linking techniques.

In this work, we report on new features in Swift that support direct
calls to Python, R, and Tcl. These features, which could easily be
extended to other scripting languages, allow Swift scripts to
orchestrate distributed execution of code written in a wide variety of
languages, currently including C, C++, Fortran, Python, R, Tcl, and
the shell.  Indeed, {\em any} external program may be called through
the shell-based technique.

The method presented here is a more approachable software development
technique for distributed-memory computing than are traditional
techniques. Using MPI, the developer could write MPI code in C and
call to an application component script (say, in Python).
In this technique, the user
would have to manage the call to the script, possibly using an
internal API specific to that language. Application data would have
to be marshaled to and from the component and among processes in a
laborious manner. The developer would have to
build significant infrastructure
to manage load balancing and other distributed computing challenges.
Alternatively, the developer could try a scripting language-specific
MPI implementation, which might ease some but not all of the described
challenges.  Additionally, that approach would limit the number of languages
that could be used; it is unlikely that communicating among MPI
processes in multiple languages would work as desired.

In our method, the developer starts with a Swift script that describes
the calls to application components in a convenient syntax.  Swift
data is passed among language-specific components implicitly as the
script progresses; no user data marshaling is required.  (MPI
messaging is used internally by the Swift/T runtime.)  Multiple
components written in different languages may be brought
together. Progress and load balancing are managed by the Swift
runtime. Overall, the approach provides a coherent programming model,
allows for compatibility among multiple languages, provides high 
scalability, and is compatible with advanced architectures such as the
Cray XE6 and Blue Gene/Q.

The rest of this paper is as follows.  In
Section~\ref{section:architecture} we describe the architecture of
Swift/T, and in Section~\ref{section:interfaces} we describe the
interlanguage features that are the focus of this paper.  In
Section~\ref{section:conclusion} we offer concluding remarks and a
glimpse of future work.

\section{Architecture}
\label{section:architecture}

We next provide some background on the Swift language, describe
the Swift/T architecture, and discuss how Swift/T
calls application components.

\subsection{Swift language}

Swift is a scripting language with C-like syntax, with pervasive,
automatic concurrency built into the language.
Concurrency is achieved through dataflow processing, in which progress
depends on the availability of input data, not statement
ordering. For example, in the code fragment

\begin{center}
\scriptsize

\begin{tabular}{r|l}
 1 & {\tt int\ x;                                 } \\
 2 & {\tt x\ =\ f(3);                             } \\
 3 & {\tt int\ y1\ =\ g(x,1);                     } \\
 4 & {\tt int\ y2\ =\ g(x,2);                     } \\

\end{tabular}
\end{center}

the declaration \texttt{int x;} creates a \textit{future} \texttt{x}.
Subsequent function calls to \texttt{g()} block until a value is
stored in \texttt{x}.  When \texttt{f()} completes, both calls to
\texttt{g()} are eligible to run concurrently on different
processors.

Massive concurrency can be achieved in Swift with relatively little
code.  For example, in the code fragment

\begin{figure}[h]
  \begin{center}
    \includegraphics[scale=0.5]{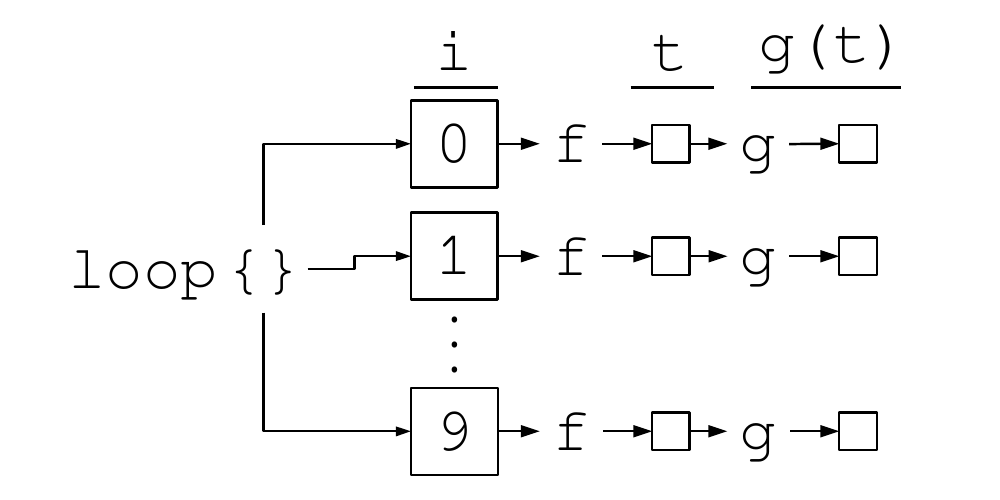}
  \end{center}
  \caption{Diagram of implicit dataflow of Swift loop.
    \label{figure:dataflow}}
\end{figure}

\begin{center}
\scriptsize

\begin{tabular}{r|l}
 1 & {\tt foreach\ i\ in\ [0:9]\ \{               } \\
 2 & {\tt \ \ int\ t\ =\ f(i);\                   } \\
 3 & {\tt \ \ if\ (g(t)\ ==\ 0)\ \{\ printf("g(\%i)==0",\ t);\ \}} \\
 4 & {\tt \}                                      } \\

\end{tabular}
\end{center}

the \texttt{foreach} loop executes each loop body for a unique value
of \texttt{i} from 0...9 concurrently.  Each execution of \texttt{f()}
may be run concurrently, but each \texttt{g(t)} is blocked on the
corresponding \texttt{f(t)}.  The code implies the
dataflow dependencies shown in \figurename~\ref{figure:dataflow},
where several parallel pipelines of tasks are
present.  Swift will construct and execute these pipelines
in parallel on any available resources.

In
the Swift model, bulk user computation is performed in \emph{leaf
  tasks}: user code outside of Swift, such as libraries or external
programs.  These are load-balanced between available processors by
dispatching tasks on demand\ketanote{how is this load-balancing?}. If
\texttt{f()} and \texttt{g()} are compute-intensive functions with
varying runtimes, the asynchronous, load-balanced Swift model is an
excellent fit.

\subsection{Swift/T runtime}

Swift/T~\cite{Swift_T_2013} is a reimplementation of the
Swift/K~\cite{Swift_2011} framework for high-performance computing.

Swift/K excels at distributed, grid, and cloud computing, and offers
wide-ranging support for schedulers (PBS, LSF, SLURM, SGE, Condor,
Cobalt, SSH) and data transfer, fault tolerance, and other features
useful for that environment. K~indicates that the language is
implemented atop the Karajan workflow engine.

Swift/T is designed for high-performance computing at the largest
scale.  T indicates that the key features are implemented by the
Turbine dataflow engine~\cite{Turbine_2013}. In this implementation of
Swift, the Swift script is translated into a runtime framework based
on the MPI-based Asynchronous Dynamic Load Balancer
(ADLB)~\cite{ADLB_2010} and Turbine libraries, which evaluate Swift
semantics in a distributed manner (no bottleneck).  Thus, at run time,
Swift/T programs are MPI programs.

\begin{figure}[h]
  \begin{center}
    \includegraphics[scale=0.7]{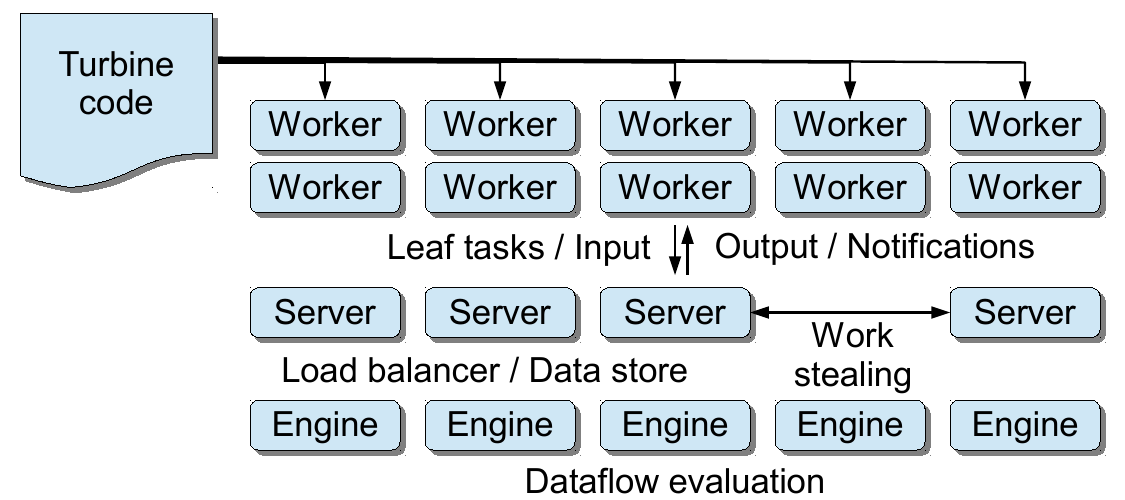}
  \end{center}
  \caption{Swift/T runtime architecture.
    \label{figure:architecture}}
\end{figure}

The Swift/T architecture is diagrammed in
\figurename~\ref{figure:architecture}.  Each MPI process operates as an
engine, ADLB server, or worker. Engines carry out Swift logic,
creating leaf tasks for execution. ADLB servers, shown as an opaque
subsystem, distribute tasks to workers, which execute user work (such
as \texttt{f()} and \texttt{g()} in our example above).  Typically the
vast majority of processes (99\%+) are designated as workers.
The engine and server processes are called \emph{control processes}
and collectively orchestrate script execution.

\section{Swift interfaces to various languages}
\label{section:interfaces}

Swift/T has multiple new methods not reported previously for calling
to user code .  In this section, we consider these in detail.

\subsection{Calling Tcl}

The Swift/T compiler (STC) translates user Swift code to a
representation (Turbine code) that uses the Turbine, ADLB, MPI, and
user libraries, all of which are written in C.  While STC could
generate C code, we desired a compiler target with the following
properties: 1) A straightforward way to ship code fragments through
ADLB for load balancing and evaluation elsewhere, 2) A textual, easily
readable format, and 3) A runtime that did not require the user to run
the C compiler in order to avoid complexities on advanced systems.
Thus, we chose Tcl to represent Turbine code, and made use of the ease
of calling C from Tcl in order to bind the system together.

Since Swift/T runs on Tcl, calling from Swift to Tcl is the most
advanced interlanguage feature in Swift/T. Consider the Swift code
fragment

\begin{center}
\scriptsize

\begin{tabular}{r|l}
 1 & {\tt (int\ o)\ f(int\ i,\ int\ j)            } \\
 2 & {\tt "my\_package"\ "1.0"                    } \\
 3 & {\tt [\ "set\ <<o>>\ [\ f\ <<i>>\ <<j>>\ ]"\ ];} \\
 4 & {\tt ...                                     } \\
 5 & {\tt int\ x\ =\ f(2,\ 3);                    } \\

\end{tabular}
\end{center}

In this code, Tcl procedure \texttt{f} is made available to Swift with
the given signature.  When inputs \texttt{i} and \texttt{j} are
available, the Tcl code (line 3) is executed.  The Tcl package
\texttt{my\_package~1.0} is loaded on the assumption that \texttt{f}
will be found in that package.  The Swift/T runtime supports user
additions to \texttt{TCLLIBPATH} so that arbitrary Tcl code may be
attached to a Swift/T run.

Interlanguage operation is supported by 1) inserting dataflow
semantics to the interface between Swift/T and Tcl and 2) automatic
type conversion. The Tcl code on line 3 cannot execute until inputs
\texttt{i} and \texttt{j} are set and transmitted to the worker on
which the code will be executed, and storage for output \texttt{o} has
been allocated. This code is automatically inserted into the compiler
output by STC and is hidden from the user (by default).  The
programmer provides a template for the Tcl code. Double angle brackets
\texttt{<<$\cdot$>>} indicate that a variable should appear in that
location. Swift/T variables are automatically converted to the
appropriate Tcl types, which are oriented toward string
representations.

The ease of interlanguage operation here offers multiple beneficial
features to Swift/T development and application users. First, the ease
of exposing simple Tcl snippets to Swift allowed for the rapid
development of Swift builtins such as \texttt{printf()},
\texttt{strcat()}, etc.  Many Tcl features can easily be brought into
Swift this way. Second, Swift users often express a desire to mix
dataflow programming with short fragments of imperative code.  This is
easily done by extending the Tcl fragment on line 3 to a multiline
script snippet, using the Swift multiline string syntax.  Certain
arithmetical or string expressions may be easier to perform in Tcl
than in Swift, especially for experienced Tcl or shell programmers.
Third, existing components built in Tcl can easily be brought into
Swift by using Swift support for Tcl packages.  Fourth, the strength
of Tcl support for calling native code is easily brought into Swift as
well, as described in the following subsection.

\subsection{Calling native code}

A primary goal of Swift/T is to speed the development process for
scaling existing codes in compiled languages (C, C++, Fortran) to
high-performance systems.  Thus, good support for calling these
languages is paramount. Tcl provides good support for calling native
code, and good tools such as SWIG are available. This approach has
demonstrated the ability to successfully call native code in many
applications, including applications that may be expressed as MPI
libraries~\cite{Swift-MPI_2013}.

\begin{figure}
 \begin{center}
  \includegraphics[scale=0.75]{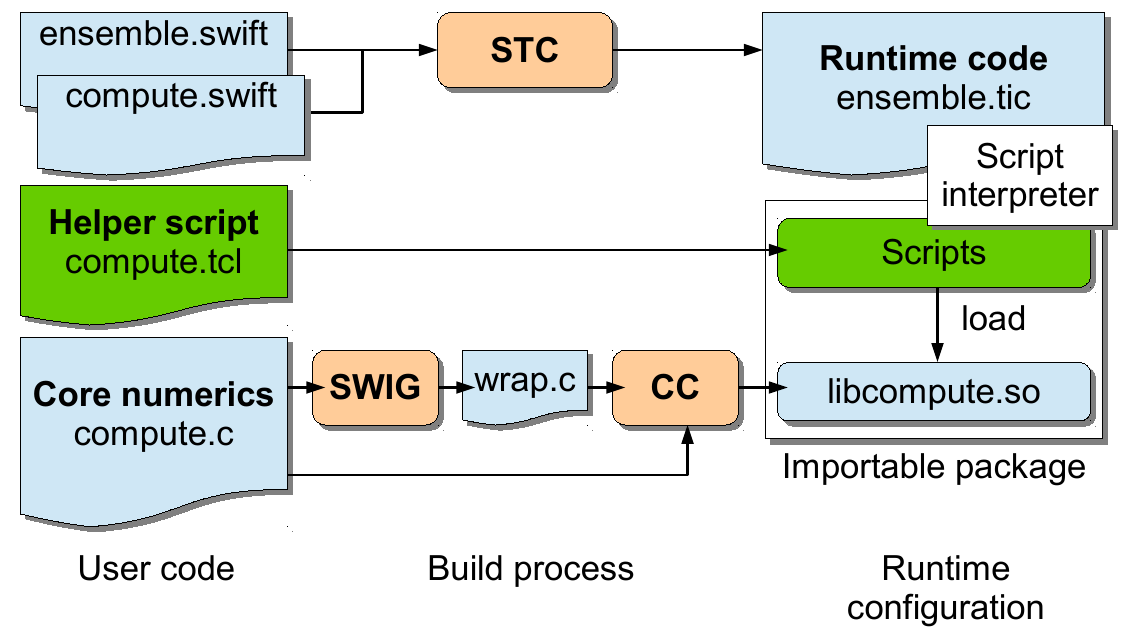}
  \caption{SWIG providing Tcl bindings for C functions callable from
    Swift/T.
  \label{figure:swig}}
  \end{center}
\end{figure}

In order to call into an existing native code program from Swift, the
following steps must be followed.  First, the user identifies the key
functions to be called.  Simple types (numbers, strings) must be used
to ensure compatibility with Swift.  Second, the program is compiled
as a loadable library - any use of \texttt{main()} must be removed
through conditional compilation.  Third, the library headers are
processed by SWIG to generate Tcl bindings for the C/C++ functions; in
the case of Fortran, a C++-formatted header is first created with
FortWrap~\cite{FortWrap_WWW}, then processed by SWIG.  Fourth, the
user writes Swift bindings for the generated Tcl bindings as described
in the previous subsection. Fifth, a Tcl package is constructed
containing the native code library and any additional Tcl scripts that
the user desires to include. Figure~\ref{figure:swig} illustrates the
process of binding a C code with Tcl using SWIG. The functions in
object \texttt{afunc.o} become callable from within Swift/T code.

The interlanguage complications here are more challenging than that in
the Tcl case because more language considerations must be taken into
account.  Our approach has been to delegate complexities and
conventions to SWIG, since it is a general-purpose tool (i.e.,
learning SWIG has broader utility than learning a Swift-specific
tool).  Thus, type conversion conventions are delegated to SWIG
conventions.

In addition to simple types, scientific users of native code languages
often desire to operate on bulk data in arrays.  The Swift approach to
these is to handle pointers to byte arrays as a novel type:
\texttt{blob} (binary large object). The Swift/T runtime handles blobs
in a similar manner to strings, but with appropriate handling for
binary data. This approach allows users to write dataflow scripts that
operate on C-formatted strings and arrays, contiguous binary data
structures, and even multidimensional Fortran arrays.

SWIG supports operations functions that consume and produce pointers
as represented by Tcl variables. Thus, Swift/T provides a small
library called \texttt{blobutils} to handle transmission
\katznote{translation?} of the Swift/T blob type to raw pointers
compatible with SWIG. Type conversion routines are provided to handle
many common cases.  For example, SWIG will not automatically convert
\texttt{void$\ast$} to \texttt{double$\ast$}-- \texttt{blobutils}
provides tools to handle the simple but myriad interlanguage
complexities found when operating on binary data.

\subsection{Calling Python or R}

As described above, many modern scientific applications have key
components or interfaces built in Python, R, or other high-level
languages.  Previous workflow programming systems call external
languages by executing the external interpreter executables.  This
strategy is undesirable for Swift/T, however, because at large scale
the filesystem overheads are unacceptable. Additionally, on
specialized supercomputers such as the Blue Gene/Q, launching external
programs is not possible at all.

Our approach, based in Swift/T, treats the external interpreters for
Python and R as native code libraries.  Thus, the complexity of
calling them is reduced to the complexity of calling a C library from
Swift/T, which was addressed in the previous section.  First, a Tcl
extension for each language was constructed.  (These could conceivably
be reused by non-Swift developers who simply desire to call Python or
R from Tcl.)  Then, a Swift/T leaf function was written that evaluates
fragments of code.  Users interact only with the high-level Swift/T
leaf function, greatly reducing complexity.

In the Swift model, each task is started without state; only the
well-defined Swift inputs are available.  When calling into an
external interpreter, however, old state from the previous task could
be available and cause confusion or debugging issues (this is not a
security issue, since all of this state is inside the Swift/T MPI
run).  One approach is to finalize the interpreter at the end of each
task and reinitialize it when the next task is started, thus clearing
any state.  This approach raises concerns about performance and
possible resource leaks.  Thus, we provide options to either retain
the interpreter or reinitialize it.  (Old interpreter state can also be
used to store useful data if the programmer is careful.)

\section{Conclusion}
\label{section:conclusion}

Modern scientific application development is trending toward greater
software complexity and more demanding performance requirements.
These applications blend structured and unstructured computing
patterns, features for distributed and parallel computing, and the use
of specialized libraries for everything from numerics to I/O.
For continued progress in scientific computing, tools must be developed
and adopted that enable rapid prototyping and development of complex,
large scale applications.

In this work, we provided a broad overview of relevant scientific
computing applications that combine computing patterns and use
multiple languages.  We described the Swift/T system for
high-performance computing, highlighted its new features to support
scripting languages such as Python and R, and showed how these can be
combined to solve numerical problems.  We also described the rich
shell interface retained and extended from Swift/K.  We described our
use of embedded script interpreters, making interlanguage programming
relatively easy while remaining compatible with systems having
restricted OS functionality such as the IBM Blue Gene/Q.  Additionally,
we showed how the many small file problem common in scripted solutions
can be addressed with our static packages.

In future work, we intend to improve support for external languages by
improving support for automatically translating more complex data
types.  Future applications are sure to challenge the current
performance envelope, and we will improve and apply our techniques to
solve bigger problems with more advanced tools on the largest scale
machines.

\section*{Acknowledgments}

This material was based upon work supported by the U.S. Dept. of
Energy, Office of Science, Office of Advanced Scientific Computing
Research, under Contract DE-AC02-06CH11357.  Work by Katz was
supported by the National Science Foundation while working at the
Foundation.  Any opinion, finding, and conclusions or recommendations
expressed in this material are those of the author(s) and do not
necessarily reflect the views of the National Science Foundation.

\bibliographystyle{IEEEtran}
\bibliography{interlang}

\end{document}